\documentclass[a4paper,12pt]{article}
\usepackage{amssymb,latexsym,amsmath}

\usepackage{fullpage}

\usepackage{mathrsfs}
\usepackage{bbold}

\usepackage[LGR,T1]{fontenc}
\usepackage[pdftex]{graphicx}
\usepackage[table,dvipsnames]{xcolor}
\usepackage{multirow}
\usepackage{rotating}
\usepackage{graphicx}
\usepackage{ulem}
\usepackage[pdftex]{hyperref}
\usepackage{ytableau}

\makeatletter \@addtoreset{equation}{section}
\makeatother

\usepackage[shadow,textwidth=2.7cm]{todonotes}
\usepackage{ifthen}
\reversemarginpar

\DeclareMathAlphabet{\mathgtt}{LGR}{cmtt}{m}{n}

\allowdisplaybreaks[1]

\begin{document}

\begin{titlepage}
	\thispagestyle{empty}
	
	\vspace{35pt}
	
\begin{center}
	
{ \LARGE{\bf  New Torsional Deformations of  \\\vskip 5mm Locally AdS$_3$ Space }}

\vspace{50pt}
		
{L.~Andrianopoli$^{1,2}$, B.~L.~Cerchiai$^{3,2}$, R.~Noris$^{4}$, L.~Ravera$^{1,2}$, M.~Trigiante$^{1,2}$, J.~Zanelli$^{5,6}$}
		
\vspace{25pt}
$^1${Politecnico di Torino, Corso Duca degli Abruzzi 24, 10129 Torino, Italy}

\vspace{15pt}
$^2${INFN, Sezione di Torino, Via P. Giuria 1, 10125 Torino, Italy}

\vspace{15pt}
$^3${IMATI Milano -- CNR, Via A. Corti 12, 20133 Milano, Italy}

\vspace{15pt}
$^4${CEICO, Institute of Physics of the Czech Academy of Sciences, Na Slovance 2, 182 21 Prague 8, Czech Republic}
		
\vspace{15pt}
$^5${Centro de Estudios Cient\'{\i}ficos (CECs), Av. Arturo Prat 514, Valdivia, Chile}
		
\vspace{15pt}
$^6${Universidad San Sebasti\'an, General Lagos 1163, Valdivia, Chile}

\end{center}
	
\vspace{10pt}
	
	\begin{abstract}
We consider general torsion components in three-dimensional Einstein-Cartan gravity, providing a geometrical interpretation for matter, and find new solutions of the corresponding equations for the Riemann curvature and torsion. These geometries involve a peculiar interplay between the vector $(\beta_i)$ and the singlet $(\tau)$ irreducible components of the torsion which, under general conditions, feature a formal analogy with the equation for a Beltrami fluid.\\ Interestingly, we find that the local AdS$_3$ geometry is now deformed by effect  of the "Beltrami-torsion" $\beta_i$. 
Some of these new solutions describe deformations of the BTZ black hole due to the presence of torsion. The latter acts as a geometric flux which, in some cases, removes the causal singularity. 

\end{abstract}

\end{titlepage}

\baselineskip 6 mm

\date{\today}


\tableofcontents

\section{Introduction} 
Over the years, three-dimensional gravity with negative cosmological constant has attracted considerable interest in view of its many remarkable features \cite{Brown:1986nw,Achucarro:1986uwr,Witten:1988hc,Achucarro:1989gm,Coussaert:1995zp,Witten:2007kt}. Among these is the fact that it can be formulated in the language of gauge theories, as a Chern-Simons action for an SO$(2,2)$ connection \cite{Achucarro:1986uwr,Witten:1988hc}. Moreover, despite the absence of local degrees of freedom, including gravitational waves, three-dimensional gravity exhibits non-trivial solutions, like the BTZ black hole \cite{Banados:1992wn,Banados:1992gq}. This solution, in particular, features a singularity in the causal structure of spacetime, different from the curvature singularity of four-dimensional black holes, and it is a natural candidate for an exactly solvable model of quantum black holes \cite{Witten:2007kt}. Other three-dimensional black holes, and their relation with the BTZ solution, were more recently discovered in \cite{Anninos:2008fx}, in the framework of topologically massive gravity. These features have sparked considerable interest in the gravity and supergravity community, in particular in relation to the AdS/CFT correspondence \cite{Maldacena:1997re} [see for example \cite{Ryu:2006bv,Nishioka:2009un,deBoer:2013vca}].\\\newline
In this paper we consider the spacetime torsion in the presence of a flat connection as a way of introducing, in the Einstein-Cartan formalism, a coupling to matter. We present novel three-dimensional solutions, while discussing their properties and their relation to the BTZ black hole.\\
A well-known example in this context is given by a constant, totally antisymmetric torsion contribution 
\begin{align}\label{adstor}
T^i[e,\omega]=de^i+\omega^i{}_j\wedge e^j=\tau\epsilon^{ijk}e_j\wedge e_k\,,\qquad \tau=\mathrm{constant}\,.
\end{align}
Here $i,j,\ldots=0,1,2$ are rigid indices in the vector representation of the local Lorentz group SO$(1,2)$, while $e^i$ and $\omega^{ij}$ are the vielbein and Lorentz connection 1-forms, respectively. In this case, the torsion $\tau$ can be traded for a negative cosmological constant: this can be shown by redefining the spin connection as $\omega^{ij}=\mathring\omega^{ij}+k^{ij}$, in terms of a torsionless, non-flat spin connection $\mathring\omega^{ij}$ and a contorsion tensor $k^{ij}$ \cite{Alvarez:2014uda}. The deformation induced by the torsion can be captured by the first-order action
\begin{align}
S[e,\mathring\omega]={-\frac{c^4}{16\pi G_N}}\int_\mathcal M \left(R^{ij}[\mathring\omega]\wedge e^k\,\epsilon_{ijk}-{\frac{\tau^2}{3}}\epsilon_{ijk}\,e^i\wedge e^j\wedge e^k\right)\,,
\end{align}
where $G_N$ is the three-dimensional Newton's constant, $\mathcal M$ is the spacetime manifold we are integrating on and $R^{ij}[\mathring\omega]=d\mathring\omega^{ij}+\mathring\omega^{i}{}_k\wedge\mathring\omega^{kj}$ is the Riemann curvature tensor.\footnote{In the following we will use natural units where $8\pi G_N=c=1$.}\\
The above special example describes locally AdS$_3$ spacetimes, which include the vacuum AdS$_3$-solution as well as BTZ black holes. Considering more general forms for $T^i[e,\omega]$, {by redefining the spin connection as above,} the torsion tensor can be seen as a geometric way of coupling matter to gravity through its contribution to the energy-momentum tensor. The latter can be derived from an action principle only in some cases and in second-order formalism.\\ \newline
A non-vanishing, non-dynamical torsion that generalises \eqref{adstor} can be conveniently written in terms of its irreducible components with respect to the Lorentz structure group SO$(1,2)$: these are the singlet $\tau$, a three-vector $\beta^i$ and a symmetric-traceless tensor $\Sigma^{ij}=\Sigma^{(ij)}$. Integrability of the torsion equation, expressed by the corresponding Bianchi identity, poses constraints on the irreducible torsion components. For the sake of concreteness, in this paper, we will focus on the $\Sigma^{ij}=0$ case and we leave a more general analysis, which includes this component, to a future endeavor. \\
In \cite{Alvarez:2011gd}, the consistency conditions on $\tau$ and $\beta\equiv \beta_i e^i$ were solved assuming $d\beta=0$. Interestingly, as we will show, these equations also allow for configurations,  $d\beta\neq0$, in which the components of $\beta$ satisfy an equation of the same general form as that of a Beltrami velocity field in fluid dynamics \cite{ABC3,Fre:2022odf}. 
Furthermore, in this more general case,  $\beta$ can be interpreted as a connection associated with the local Weyl-rescaling symmetry of the field equations. As we shall show, this can be exploited to rescale the $\tau$ component to a constant, by a local gauge choice.\\
This leads to the explicit construction of novel geometries which yield, in the $\beta\to 0$ limit, locally AdS$_3$ spacetimes. 
In particular, we obtain a class of spacetimes that describe a deformation of the BTZ black hole induced by the "Beltrami-torsion" 3-vector $\beta^i$.
In some of these solutions, the effect of  $\beta$, which can be interpreted as a geometrical flux, is to remove the causal singularity of the BTZ solution. 
\\
We present here a preliminary analysis of the geometric properties of these solutions, leaving a more detailed study thereof, as well as a discussion of their application to gauge/gravity duality, to future work.
Among the geometries considered here, we include the case of a Euclidean 3-dimensional space, where the geometric flux $\beta^i$ satisfies the same equation as the velocity of a proper Beltrami fluid.\\
Lastly, this perspective, which revolves around the spacetime torsion, is also the natural framework in \textit{Unconventional sypersymmetry} contexts, which play a relevant role in the construction of analogue supergravity models, providing a macroscopic description of the electronic properties of graphene-like materials \cite{Alvarez:2013tga,Guevara:2016rbl,Andrianopoli:2018ymh,Andrianopoli:2019sip,Alvarez:2021zhh}.
In this kind of applications, the parameter $\tau$ was given the interpretation of Semenoff mass \cite{Andrianopoli:2019sip}. In such context, the role of all the irreducible components of the torsion is currently under investigation and will be the subject of a forthcoming publication.
\\ \newline
The paper is organised as follows: in Section \ref{sec2} we introduce the notion of "Beltrami-torsion" as the general solution to the consistency conditions on the irreducible torsion components and derive the explicit form of the energy-momentum tensor induced by it. In Section \ref{sec3} we present a number of novel geometries and discuss their properties, in relation to BTZ black hole and Beltrami fluids. Finally, in Section \ref{sec4}, we discuss possible applications of the results and future developments.\\
We summarise our conventions in Appendix \ref{appA}, while Appendix \ref{appB} is devoted to the study of AdS$_3$ Killing vectors.

\section{General torsion in 2+1 dimensions}\label{sec2}

In three spacetime dimensions, the general decomposition of the torsion 2-form in irreducible representations of the Lorentz group is
\begin{equation}\label{tor1}
    T^i[e,\omega] = D{[\omega]}e^i  = \tau \,\epsilon^i_{\ jk} \, e^j\wedge e^k+ \beta \, e^i +\Sigma^{il}\,\epsilon_{l jk}e^j\wedge e^k\,,
\end{equation}
according to the branching rules for the product of the vector ($\mathbf{3}$) times the twice antisymmetric ($\mathbf{3}\,{\wedge}\,\mathbf{3}=\mathbf{3}$) SO$(1,2)$-representations 
\begin{equation}
T^i{}_{[jk]}: \quad \mathbf{3}\times \mathbf{3}= \mathbf{1} \oplus \mathbf{3}  \oplus\mathbf{5}\,.\label{reps}
\end{equation}
Here $D{[\omega]}$ denotes the Lorentz-covariant derivative with respect to a generic spin connection $\omega$; $\tau$  is the 0-form related to the totally antisymmetric irreducible representation $(T_{(1)})^i{}_{j k}\equiv\tau \,\epsilon^i{}_{jk}$ (corresponding to the singlet in \eqref{reps}); $\beta = \beta_i e^i$  is the 1-form associated with the trace part of the torsion $(T_{(3)})^i{}_{j k}\equiv\beta_{[j}\delta^i_{k]}$ (corresponding to the ${\bf 3}$ in \eqref{reps}) and the last term on the right-hand-side of \eqref{tor1}, $(T_{(5)})^i{}_{j k}\equiv\Sigma^{il}\epsilon_{l jk}$, is defined by the symmetric traceless matrix $\Sigma^{ij}$ describing the irreducible component
{\small{\tiny \begin{ytableau}
{}&{}\\
{}
\end{ytableau}}} of the torsion tensor (corresponding to the ${\bf 5}$ in \eqref{reps}).  The first and last terms in \eqref{tor1} can be collected into a single symmetric matrix
\begin{equation}
    S^{ij}=\tau \delta^{ij}+\Sigma^{ij}\,.\label{esse}
\end{equation}
Let us consider here the effects of a Weyl rescaling 
\begin{equation}
e^i \to \lambda(x) e^i\,,
\label{weyl}
\end{equation}
under which, in general, the torsion does not transform as a tensor 
$$D {[\omega]}e^i \to D {[\omega]}{e^i}'=\lambda \left(D {[\omega]}e^i + \frac 1\lambda d\lambda e^i\right)\,.
$$
Nonetheless, it is tempting to consider $\beta$ as an abelian connection, associated to a O$(1,1)$-bundle of Weyl rescalings and transforming as
\begin{eqnarray}
\beta'&=& \beta+\frac{d\lambda}\lambda\,. \label{con}
\end{eqnarray}
Indeed, let us associate the quantity 
\begin{align}
    \hat T^i[e,\omega,\beta]=D[\omega]e^i-\beta e^i=S^{il}\epsilon_{ljk}e^j\wedge e^k
\end{align}
to the torsion \eqref{tor1}. The above expression transforms covariantly $$\hat {T}^{i}[e',\omega,\beta']= \lambda \hat T^i[e,\omega,\beta]\,,$$ provided that we also require 
\begin{eqnarray}
S^{\prime \,ij}&=&\frac {1}{\lambda}\,S^{ij}\,\label{s'}
\end{eqnarray}
to hold. Under these circumstances, $\hat T[e,\omega,\beta]^i$ becomes a O$(1,1)$-tensor. Quantities transforming tensorially under Weyl rescalings can then be obtained by defining generalised (Lorentz and O$(1,1)$) covariant derivatives
\begin{align}
\hat {D}[\omega,\beta]\Phi^{(p)}\equiv {D}[\omega]\Phi^{(p)}-p\,\beta \wedge \Phi^{(p)}\,,
\end{align}
$\Phi^{(p)}$ denoting any of the fields/parameters of the theory and $p$ the corresponding O$(1,1)$-weight. In particular, the dreibein $e^i$ has O$(1,1)$-weight $p=1$, whereas $S^{ij}$ is a section of the O$(1,1)$-bundle with weight $p=-1$.\\ 
The introduction of such a symmetry implies that it is always possible to rescale the value of the singlet function $\tau$ to a constant.\\\newline
In the following, we shall consider a flat Lorentz-connection $\omega$ and restrict to the torsion components $\tau,\,\beta$ only, leaving a more general study, which includes the $\Sigma^{ij}$ component, to future work. The relevant field-strengths will then have the following form:
\begin{align}\label{R}
    R^{ij}[\omega]&=0\,,\\
    T^i[e,\omega] &= \tau \epsilon^{ijk} e_j\wedge e_k+ \beta \wedge e^i\,.\label{T}
\end{align}

\subsection{The torsion Bianchi identity}
The torsion equation \eqref{T} must preserve the nilpotency of the differential operator $d$. This yields integrability conditions involving $\tau$ and $\beta$. The Bianchi identity for the torsion reads:
\begin{align}\label{btor0}
    DT^i[e,\omega]&=R^{ij}[\omega]\wedge e_j=0\,,
\end{align}
{where we have used \eqref{R}.\footnote{The same condition would hold for any Einstein manifold, where the Riemann tensor is $R^{ij}{[\omega]}\propto e^i\wedge e^j$.} The above equation implies, upon using \eqref{T}, the following condition on the irreducible components $\beta$ and $\tau$:
\begin{align}\label{btor}
  0&=d\beta\wedge e^i+\epsilon^{ijk}(d\tau+\beta\tau)\wedge e_j\wedge e_k\,.
\end{align}}
Equation \eqref{btor} can be rewritten, using differential forms, as
\begin{align}\label{integrability}
    \star d\beta=-2(d\tau+\beta\tau)\,,
\end{align}
where the Hodge star has been defined in Appendix \ref{appA}. The rigid component expression of the above equation is 
\begin{align}
    D^{[i}[\mathring\omega]\beta^{j]}+\epsilon^{ijk}(\partial_k\tau+\beta_k\tau)=0\,,\label{cancem}
    \end{align}
while, in terms of curved spacetime indices, equation \eqref{integrability} can equivalently be expressed as 
\begin{equation}
   \frac 12 \sqrt{g}\,{\epsilon_\mu}^{\nu\rho}\partial_\nu\beta_\rho =-(\partial_\mu\tau+\beta_\mu\tau)\,.
\end{equation}
In the $\tau\neq0$ case, if we require covariance of the torsion under Weyl rescalings, \eqref{weyl}, we can suitably fix the O$(1,1)$ symmetry, in such a way that the totally antisymmetric component of the torsion becomes constant. Indeed, taking $\lambda=\frac{\tau}{\alpha}\neq0$, with $\alpha$ constant, yields
\begin{align}
    \tau'=\alpha\,,   \qquad\beta'=\beta+d\ln\left(\frac{\tau}{\alpha}\right)\,,\qquad e'\,{}^i=\frac{\tau}{\alpha}e^i\,.
\end{align}
In this case, the integrability equation, which is covariant with respect to O$(1,1)$ transformations, reduces to 
\begin{align}\label{ArnoldBeltrami}
    \star' d\beta'=-2\beta'\tau'\,.
\end{align}
The above equation interestingly leads to a Proca-like equation for a massive field $\beta'$, which reads
\begin{align}
    \Box\beta'=(\star' d\star'd)\beta'=4\beta'\tau'\,^2\,.
\end{align}
The case in which $\beta'$ is closed, and actually vanishing, implying that $\beta$ is exact, has already been discussed in the literature \cite{Alvarez:2011gd}. In the $d\beta'\neq0$ case, \eqref{ArnoldBeltrami} is reminiscent of the {equation defining a Beltrami flow in  fluid-dynamics} and will lead to the novel and interesting geometries, that we discuss in the next Section.  For this reason, we will generally refer to $\beta$, satisfying \eqref{ArnoldBeltrami}, as "Beltrami-torsion". \\
Finally, let us notice that this is a special case of the generalised monopole equation discussed in \cite{Jones:1985pla} (see also \cite{Klemm:2020gfm}), though in a different context.

\subsection{Energy-momentum tensor from torsion}
As mentioned in the Introduction, since the explicit form of the torsion depends on the choice of spin connection, a generic torsion can always be set to zero by redefining the spin connection in terms of a torsionless one, $\mathring\omega$, plus a contorsion $k$, that is
\begin{align}
\omega^{ij}&= \mathring\omega^{ij}+ k^{ij}\,,\\
R^{ij}[\omega]&=  R^{ij}[\mathring\omega]+D[\mathring \omega] k^{ij}+ k^i{}_k\wedge k^{kj}\,.
\end{align} 
By doing so,  the resulting Riemann tensor receives a shift, {that can be regarded as (minus) the energy-momentum contribution of some sort of matter and/or flux to the Einstein equations. \\
In fact, in the case at hand, the contorsion tensor explicitly reads
\begin{align}
    k^{ij}=-\tau\epsilon^{ijk}e_k+2\beta^{[i}e^{j]}\,,
\end{align}
and \eqref{R} and \eqref{T} can be rewritten in terms of the torsionless spin connection as
\begin{align}
    R^{ij}[\mathring\omega]&=-\frac12\epsilon^{ijk}\epsilon^{lpq} \mathcal T_{kl} e_p\wedge e_q\,, \label{Rtorsionless}\\
    T^i[e,\mathring\omega] &=0\,.\label{Ttorsionless}
\end{align}
Here the tensor $\mathcal T^{ij}$ is defined as
\begin{align}\label{EMTgeneral}
    \mathcal T^{ij}=\left(-\tau^2\eta^{ij}-\beta^i\beta^j-(\partial_k\tau+\beta_k\tau)\epsilon^{ijk}-D^i[\mathring\omega]\beta^j+D^k[\mathring\omega]\beta_k\eta^{ij}\right)\,,
\end{align}
and in principle has both symmetric and antisymmetric parts. However, in virtue of \eqref{cancem}, the antisymmetric part automatically vanishes, leaving a symmetric tensor, which reads 
\begin{align}
    \mathcal T^{ij}=\left(-\tau^2\eta^{ij}-\beta^i\beta^j-D^{(i}[\mathring\omega]\beta^{j)}+D^k[\mathring\omega]\beta_k\eta^{ij}\right)\,.
\end{align}
The symmetric tensor $\mathcal T^{ij}$ is in fact the energy-momentum tensor of the theory: by expanding the Riemann tensor in the vielbein basis and by taking appropriate traces, one can build the Einstein's tensor, which precisely reads
\begin{align}
    \mathcal  G_{ij}=\mathcal R_{ij}[\mathring\omega]-\frac12\eta_{ij}R[\mathring\omega]=\mathcal T_{ij}\,,
\end{align}
where $\mathcal R_{ij}[\mathring\omega]$ indicates the Ricci tensor, whereas $R[\mathring\omega]$ is the corresponding Ricci scalar. {At last, observe that the right-hand side is automatically covariantly conserved, in virtue of the Riemann curvature Bianchi identities, in this torsionless spin connection frame. Indeed, $D[\mathring\omega]R^{ij}[\mathring\omega]=0$ precisely implies $D_i[\mathring\omega]\mathcal T^{ij}=0$. This is a quadratic equation on the contorsion tensor condition, but it exactly coincides with the linear integrability condition \eqref{integrability}, manifestly showing the advantages of working in the torsionful framework.}

\section{The solutions}\label{sec3}
In this Section, we present new solutions of equations \eqref{R} and \eqref{T} and we discuss their geometric features. 

\subsection{Spacelike fibration solutions}
Let us consider a constant, completely antisymmetric torsion component $\tau$ and a Beltrami torsion $\beta$ proportional to $e^2$, defining a spacelike fibration.\\
A one parameter family of solutions is given by 
\begin{align}\nonumber\label{sol}
    e^0&=\frac{1}{2 \tau}dt\,,\\ \nonumber    
    e^1&=\frac{1}{2\tau}{{dx} \left( \cos \left(\sqrt{1-\zeta^2} t\right)\right)}\,,\\ \nonumber
    e^2&=\frac{1}{2\tau}\left({dz}+dx\left({\,\frac{  \sin \left(\sqrt{1-\zeta^2} t\right)}{\sqrt{1-\zeta^2}}}\right)\right)\,,\\
    \beta&=2\tau\zeta e^2\,, 
\end{align}
whose associated metric tensor reads
\begin{align}
    ds^2=\frac{{dt}^2-{dz}^2-{dx}^2 \left(\cos ^2\left(\sqrt{1-\zeta ^2} t\right)+\frac{\sin ^2\left(\sqrt{1-\zeta ^2} t\right)}{1-\zeta ^2}\right)-2 {dx}\, {dz}\frac{ \sin \left(\sqrt{1-\zeta ^2} t\right)}{\sqrt{1-\zeta ^2}}}{4 \tau^2}\,.
\end{align}
These expressions are written in terms of local coordinates $(t,x,z)$, each covering the entire real line. As the metric components depend only on the coordinate $t$, one has the following non-timelike Killing vectors:
\begin{align}
    K_x=\frac{\partial}{\partial x}\,,\qquad K_z=\frac{\partial}{\partial z}\,,
\end{align}
which one can show being the only admissible ones (together with their linear combinations) for generic values of the parameter $\zeta$.\\ 
The corresponding torsionless Riemann tensor reads
\begin{align}
    R^{01}[\mathring\omega]&=(1-4\zeta^2)\tau^2 e^0\wedge e^1\,, \qquad  R^{02}[\mathring\omega]=\tau^2 e^0\wedge e^2\,,\qquad  R^{12}[\mathring\omega]=\tau^2 e^1\wedge e^2\,,
\end{align}
whereas the Ricci scalar is
\begin{align}
    R[\mathring\omega]&=6\tau^2-8\zeta^2\tau^2\,.
\end{align}
Notice that the parameter $\zeta\in \mathbb R$ controls the strength of the deformation with respect to an AdS$_3$ space.\footnote{Notice that the value $\zeta=1$ can be safely included by supplementing the function $\frac{  \sin \left(\sqrt{1-\zeta ^2} t\right)}{\sqrt{1-\zeta ^2}}$ with the value $t$, as $\zeta=1$. In this way the vielbein becomes continuous and twice differentiable in $\zeta=1$.} In particular, the effect of the spacelike fibration is that of introducing an AdS$_3$ \textit{squashing} along the $01$ component of the Riemann tensor. Most interestingly, observe that each Lorentz component of this tensor is still proportional to the very same combination of vielbein and that the resulting spacetime is still an Einstein manifold. \\
When the Beltrami torsion vanishes, the metric reduces to 
\begin{align}
    ds^2\Big|_{\zeta\to0}=\frac{{dt}^2-{dx}^2-{dz}^2-2{dx}\, {dz} \sin (t)}{4 \tau^2}\,,
\end{align}
which can be obtained from the following embedding $(t,x,z)\mapsto (X^M)\in \mathbb R^{2,2}$ 
\begin{align}\nonumber
    X^0&=\frac{1}{\tau}{\cos \left(\frac{1}{2} \left(t-\frac{\pi }{2}\right)\right) \cosh \left(\frac{x+z }{2}\right)}\,,\\
    X^1&=\frac{1}{\tau}{\cos \left(\frac{1}{2} \left(t-\frac{\pi }{2}\right)\right) \sinh \left(\frac{x+z }{2}\right)}\,,\\ \nonumber
    X^2&=\frac{1}{\tau}{\sin \left(\frac{1}{2} \left(t-\frac{\pi }{2}\right)\right) \sinh \left(\frac{z -x}{2}\right)}\,,\\ \nonumber
    X^3&=\frac{1}{\tau}{\sin \left(\frac{1}{2} \left(t-\frac{\pi }{2}\right)\right) \cosh \left(\frac{z -x}{2}\right)}\,,
\end{align}
satisfying 
\begin{align}
    (X^0)^2-(X^1)^2-(X^2)^2+ (X^3)^2=\frac{1}{\tau^2}\,.
\end{align}
The above equations are well defined for all values of $t,x,z$ and they show that the coordinates $(t,x,z)$, in the $\zeta\to0$ limit, only describe a local patch of AdS$_3$, as $0<(X^0)^2-(X^1)^2<\frac{1}{\tau^2}$, instead of covering the entire real line, as in the global coordinate case.\\\newline
The torsionless spin connection is
\begin{align}\label{spinconnection}
    \mathring\omega^{01}&=2\tau\sqrt{1-\zeta^2}\,\tan(t\sqrt{1-\zeta^2})\,e^1-\tau e^2\,,\qquad 
    \mathring\omega^{02}=-\tau e^1\,,\qquad
    \mathring\omega^{12}=-\tau e^0
\end{align}
and the covariantly conserved energy-momentum tensor associated to this configuration reads
\begin{align}
    \mathcal T^{ij}=\left(-\tau^2\eta^{ij}-4\tau^2\zeta^2\delta^i_2\delta^j_2\right)\,.
\end{align}
One can immediately see that this correctly 
reduces to a cosmological constant contribution in the vanishing $\zeta$ limit and that it is symmetric. 

\subsubsection{Relation to BTZ black holes}
{Let us analyse here the above-described solution in relation to the well-known BTZ black hole. To this end, we will consider two appropriate coordinate patches and study the associated Killing vectors. We will then discuss the possibility of performing quotients with respect to spacelike Killing vectors and the consequences on the global geometry.}\\
Let us start by defining a convenient parameter $\xi=\sqrt{1-\zeta^2}$. By doing so, we are restricting to the $\zeta\in[0,1]$ range of admissible values, which in turn correspond to $\xi\in[0,1]$. The vielbein can then be rewritten as
\begin{align}\nonumber
    e^0&=\frac{1}{2 \tau}dt\,,\quad    e^1=\frac{1}{2\tau}{{dx} \left( \cos \left(\xi t\right)\right)}\,, \quad e^2=\frac{1}{2\tau}\left({dz}+dx\left({\,\frac{  \sin \left(\xi t\right)}{\xi}}\right)\right)\,. 
\end{align}

\paragraph{No singularity case:} \ \\
In a different coordinate patch, parametrised by $(\text{t},r,\phi)$, the solution \eqref{sol} can be rewritten, for $\xi\in (0,1]$ as 
\begin{align}\nonumber
    e^0&=\frac{\tau d\text{t}-d\text{$\phi $} }{(r_+-r_-)}\sqrt{(r^2-r_+^2) (r^2-r_-^2) }\,,\\ \nonumber
    e^1&=-\frac{{dr}\, r}{\xi  \tau  \sqrt{(r^2-r_+^2) (r^2-r_-^2)}}\,,\\
    e^2&=-\frac{\xi  (\tau d\text{t}+d\text{$\phi $}) (r_+-r_-)^2-(\tau d\text{t}-d\text{$\phi $}) \left(2 r^2-r_+^2-r_-^2\right)}{2 \xi (r_+-r_-)}\,,
\end{align}
with $r_+>r_->0$. Here $r\in \mathbb R_+\setminus \{ r_\pm \}$, while $t,\phi\in \mathbb R$. {While the related metric is real for any value of $r$, let us observe that the {first two components of the} vielbein become {imaginary} in the internal region $r_-<r<r_+$: for these values of  $r$, one can perform a complex Lorentz transformation \begin{align}\label{complexLorentz}
    \Lambda=\begin{pmatrix} 0& i& 0 \\
    -i&0&0\\
    0&0&-1\end{pmatrix}{\in \textrm{SO}(3;\mathbb C)}\,,
\end{align} 
which exchanges the roles of $e^0$ and $e^1$ and crucially leaves $e^2$ untouched, guaranteeing that $\beta$ is still real.}
The associated metric can be written in fibration form, like the original BTZ solution, as
\begin{equation}
    ds^2=\left(g_{\text{tt}}-\frac{g_{\text{t}\phi}^2}{g_{\phi\phi}}\right)\,d\text{t}^2+g_{rr}\,dr^2+g_{\phi\phi}\left(d\phi+\frac{g_{\text{t}\phi}}{g_{\phi\phi}}\,d\text{t}\right)^2\,,
\end{equation}
with
\begin{align}\label{metriccomp}
g_{rr}&=-\frac{r^2}{(r^2-r_+^2)(r^2-r_-^2){\tau^2}\xi^2}\,,\nonumber\\
    g_{\text{tt}}-\frac{g_{\text{t}\phi}^2}{g_{\phi\phi}}&=\frac{(r^2-r_+^2)(r^2-r_-^2)}{P(r,\xi)}\,,\nonumber\\
   g_{\phi\phi}&=-\tau^2\, P(r,\xi)\,,\nonumber\\
   g_{\text{t}\phi }&=\tau{r_+ r_-}+\frac{(2r^2-r_+^2-r_-^2)^2(1-\xi^2)\tau}{4(r_+-r_-)^2\xi^2}\,.
\end{align}
The explicit expression of the polynomial $P(r,\xi)$ 
\begin{align}
    P(r,\xi)=\frac{\left(2 r^2-r_+^2-r_-^2+\xi  (r_+-r_-)^2\right)^2-4\xi^2 \left(r^2-r_+^2\right)\left(r^2-r_-^2\right) }{4\tau^2\xi^2 (r_+-r_-)^2}
\end{align}
reduces to a simple form in the $\xi\to1$ limit $$\lim_{\xi\to1}P(r,\xi)=\frac{r^2}{\tau^2}\,.$$
Moreover let us notice that the \textit{extremal case} $r_+=r_-$ is only allowed in the $\xi=1$ case.\\ \newline
This geometry admits the following two Killing vectors, as a consequence of the fact that the metric components only depend on $r$:
\begin{align}\label{killingr}
    K_{\text t}=\frac{\partial}{\partial \text{t}}\,,\qquad K_\phi=\frac{\partial}{\partial \phi}\,.
\end{align}
{Out of the six $\mathfrak{so}(2,2)$ isometries of AdS$_3$, only the above two are preserved by the Beltrami torsion component $\beta$ and generate a $\mathfrak{so}(2)\oplus\mathfrak{so}(2)$ algebra. A detailed description of the isometry algebra of AdS$_3$, in this coordinate patch, is given in Appendix \ref{appB}.}\\
{Killing horizons} can be described as null hypersurfaces in which the norm of a Killing vector vanishes: in our case, let us consider the hypersurfaces generated by the functions $f_\pm(r)=r-r_\pm$. These functions define normal covectors $n_{\pm\mu}\propto\partial_\mu f_\pm(r)$, whose associated norm vanishes on the hypersurface itself
\begin{align}
    g^{\mu\nu}n_{\pm\mu}n_{\pm\nu}\Big|_{r=r_{\pm}}=0\,.
\end{align}
Moreover, the following linear combinations become null (and mutually orthogonal) on each hypersurface respectively:
\begin{align}
    K_\pm=\frac{r_+(1\pm\xi)+r_-(1\mp\xi)}{2\tau}K_{\text t}+\frac{r_+(1\mp\xi)+r_-(1\pm\xi)}{2}K_\phi\,.
\end{align}
We can therefore conclude that the coordinate singularities $r=r_\pm$ define Killing hypersurfaces, which, in presence of a singularity would serve as event horizons. This could also be argued by noticing that the combination $g_{\text{tt}}-\frac{g_{\text{t}\phi}^2}{g_{\phi\phi}}$ has the same zeros, $r=r_\pm$, as in the BTZ case, independently of the deforming parameter $\xi$.\\\newline
Ergospheres are another interesting type of hypersurfaces that we wish to study: these are usually defined as those loci, where the norm of the asymptotically timelike Killing vector vanishes. {In the case at hand, for a non-vanishing $\beta$, both Killing vectors in \eqref{killingr} are asymptotically spacelike. However, in the $\xi\to1$ limit, $K_{\text t}$ coincides with the asymptotically timelike Killing vector of the BTZ solution.} For this reason, let us consider the points where the norm of $K_{\text t}$ changes sign and analyse the zeros of the $g_{\text{tt}}(r)$ component of the metric. The latter reads
\begin{align}
    g_{\text{tt}}(r)=\frac{\tau ^2}{{ (r_+-r_-)^2}}\left((r^2-r_+^2)(r^2-r_-^2) -\frac{\left(2 r^2-r_+^2-r_-^2-\xi  (r_+-r_-)^2\right)^2}{4 \xi ^2}\right)\,.
\end{align}
The above expression is a quartic polynomial in $r$, which, depending on the value of $\xi$, may or may not have real, positive zeros. In the deformed solution, $\xi\neq1$, two such zeros exist, provided that 
\begin{align}
    \xi^*\leq \xi <1\,,\qquad \xi^*=\frac{r_++r_-}{\sqrt{2(r_+^2+r_-^2)}}
\end{align}
and read
\begin{align}
    r^*_{\pm}=\sqrt{\frac{\left(1-\xi ^2\right) \left(r_+^2+r_-^2\right)+\xi  (r_+-r_-)^2\pm\xi  \left(r_+^2-r_-^2\right) \sqrt{\frac{ \xi ^2 }{\xi^{*2}}-1}}{2 \left(1-\xi ^2\right)}}\,.
\end{align}
In particular, for any such value of $\xi$, we have the following inequalities:
\begin{align}
    0<r_-<r_+<r^*_{-}\leq r^*_{+}\,,
\end{align}
where the last equality holds at $\xi=\xi^*$, in which case
\begin{align}
    r^*_{+}=r^*_{-}=\sqrt{(r_+^2+r_-^2){\left(\xi^*+\frac12\right)}}\,.
\end{align}
If we keep decreasing $\xi$ past the $\xi^*$ value, thus increasing the magnitude of the deformation induced by the Beltrami torsion term, the zeros of $g_{\text{tt}}$ become complex and therefore disappear. \\
The nature of these {timelike} hypersurfaces can be better inspected by studying their properties for a small deformation: close to the undeformed solution we have
\begin{align}\nonumber\label{undeflimit}
    \lim_{\xi\to1} r^*_{+}&=+\infty\,,\\
    \lim_{\xi\to1}r^*_{-}&=\sqrt{r_+^2+r_-^2}\,,
\end{align}
while, as $\xi$ decreases, $r^*_{-}$ increases and $r^*_{+}$ decreases until they coincide and then disappear. The behaviour of $g_{\text{tt}}(r)$ and its zeros are shown in Figure \ref{gtt} and \ref{zeros} for a given choice of $r_\pm$. \\ 
\begin{figure}[ht]
\centering
\includegraphics[width=0.7\textwidth]{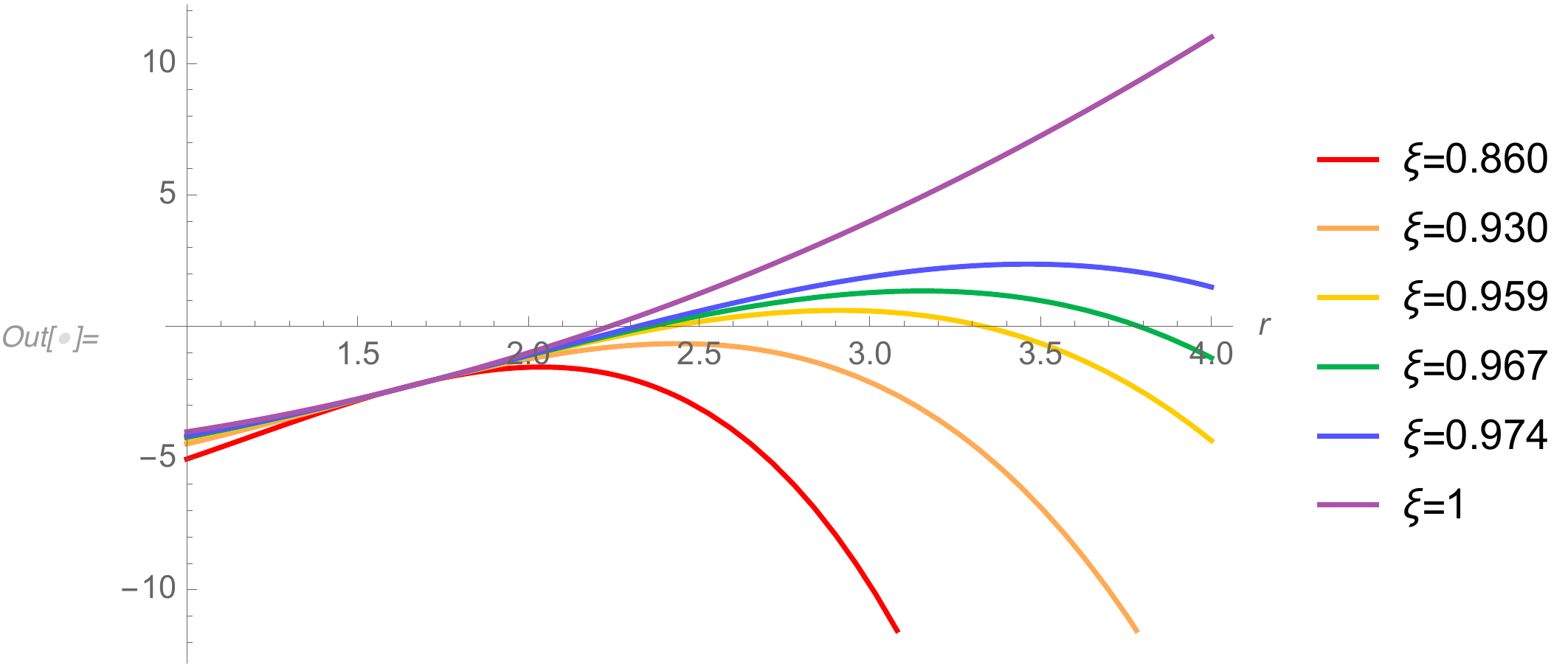}
\caption{The behaviour of $g_{\text{tt}}(r)$ is shown here for $r_-=1$, $r_+=2$ and for different values of $\xi$. In this case $\xi^*\simeq0,949$, which is consistent with the zeros that we see appearing starting from the yellow line.}\label{gtt}
\end{figure}
\begin{figure}[ht]
\centering
\includegraphics[width=0.8\textwidth]{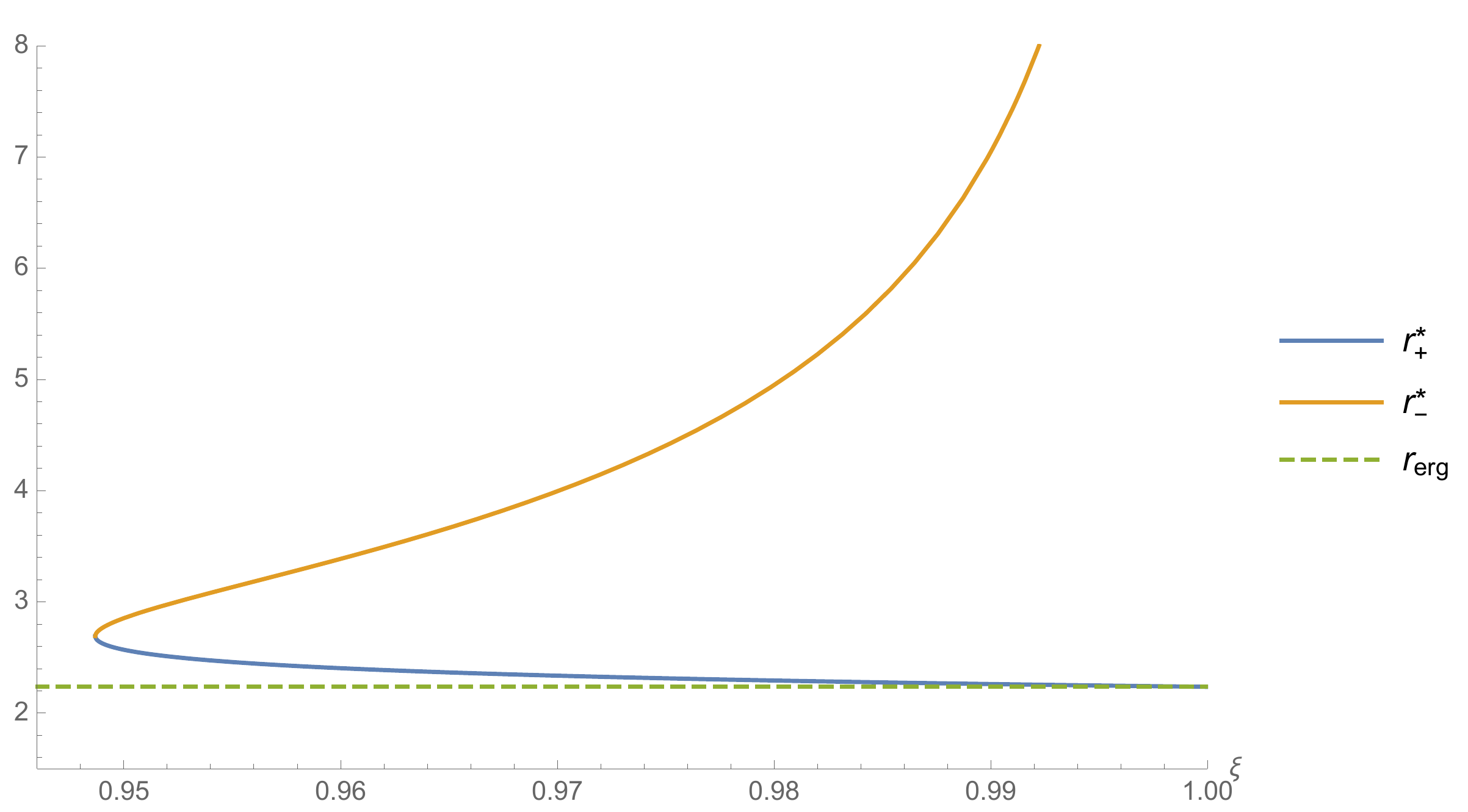}
\caption{The orange and light-blue lines represent the dependence of $r^*_{\pm}$ on the parameter $\xi$, for the values $r_-=1$, $r_+=2$. For such choice, we have $\xi^*\simeq 0,949$ and BTZ ergosphere value $r_{\mathrm{erg}}\simeq2.236$, depicted here as a green dashed line.}\label{zeros}
\end{figure}
Let us observe that, in this limit, the expression of $r_-^*$ precisely coincides with that of the ergosphere in the BTZ solution, whereas $r_+^*$ can be identified with the AdS$_3$ boundary. \\
For small deformations, $\xi>\xi^*$, the solution admits a region where the norm of the Killing vector $K_{\text t}$ is positive, $r^*_-<r<r^*_+$, indicating the presence of static, non-rotating observers between the two ergospheres. This region shrinks to zero as $\xi\leq\xi^*$, in which case all observers must rotate. 
The parameter $\xi$, which controls the strength of the torsion deformation, then also determines the width of this region, bringing the boundary of the locally AdS$_3$ spacetime to a finite $r$ value and uncovering a rotating region past infinity.\\\newline
Similarly to the BTZ solution case, let us now perform a quotient with respect to $K_\phi$, as in \cite{Banados:1992gq}: this is achieved by identifying points $p$ that belong to the same identification subgroup defined by
\begin{align}
    p\sim e^{s K_\phi}p\,, \qquad \sigma\in 2\pi\mathbb Z\,.
\end{align}
This explicitly means considering a curve $\gamma(s)$, such that $\gamma(0)=p$ and $d\gamma(s)/ds=K_\phi$ for all $s$, that is whose tangent vector is always $K_\phi$, and then identifying $\gamma(0)\sim\gamma(2\pi)$. The solution to this differential equation is $\gamma^\mu(s)=s K^\mu_\phi+p^\mu$. If $p^\mu=(t_0,x_0,\phi_0)$ and given that the Killing vector components are $K^\mu_\phi=(0,0,1)$, we have that $\gamma^\mu(s)=(t_0,x_0,\phi_0+s)$. The identification of points then implies
\begin{align}
    (t_0,x_0,\phi_0)\sim(t_0,x_0,\phi_0+2\pi)\implies \phi_0\sim\phi_0+2\pi\,,
\end{align}
which has to hold for any $p$ and therefore for any $\phi_0$. This in turn requires $\phi\sim\phi+2\pi$, which signals that the coordinate $\phi$ becomes compact.\\
Since $K_\phi$ generates isometries of the solution, the quotient space inherits the same Riemannian structure, thus resulting again, locally, in a AdS$_3$ spacetime. This identification is consistent, provided that we do not alter the causal structure, by introducing closed timelike/null curves. This problem is avoided as long as the norm of the Killing vector with respect to which we are quotienting, $K_\phi$, is negative. In the original BTZ solution, there are no problems for $r>0$, but $r=0$ defines a locus of points in which $g_{\phi\phi}(r)$ becomes zero, indicating that $K_\phi$ becomes lightlike. This locus must then be understood as a singularity of the causal structure, which is consistently hidden by two event horizons. In the case at hand, for $\xi\neq1$, the metric component $g_{\phi\phi}(r)$ never vanishes and it is always negative: we can then conclude that the presence of a non-vanishing $\beta$ removes such singularity. In the $\xi\to1$ case, the quotient space coincides with the BTZ one, whose known metric reads
\begin{align}
    ds^2_{\mathrm{BTZ}}=\frac{d\text{t}^2 \tau ^2\left(r^2-r_-^2\right) \left(r^2-r_+^2\right)}{r^2 }-\frac{{dr}^2\, r^2}{\tau ^2 \left(r^2-r_-^2\right) \left(r^2-r_+^2\right)}-{r^2}{\left(d\text{$\phi $}-\frac{d\text{t}{\tau}\, r_+r_-}{r^2}\right)^2}\,.
\end{align}
\paragraph{Singularity case:} \ \\
Let us now consider yet another patch,\footnote{This patch can be obtained from the previous one by performing the following change of coordinates:
\begin{align}
    \texttt{t}=\frac{r_+^2+r_-^2}{r_+^2-r_-^2}t-\frac{2 r_+r_-}{(r_+^2- r_-^2) \tau }\phi\,,\quad \texttt{r}=r\,,\quad \mathgtt{f}=\frac{2 r_+r_-}{(r_+^2- r_-^2) \tau }t-\frac{r_+^2+r_-^2}{r_+^2-r_-^2}\phi\,. 
\end{align}
Here we also relabeled $\texttt{r}_\pm=r_\pm$.} in which the vielbein reads
\begin{align}\nonumber
    e^0&=\frac{\tau d\texttt{t}+d\mathgtt{f}}{(\texttt{r}_++\texttt{r}_-)}\,\sqrt{(\texttt{r}^2-\texttt{r}_+^2)(\texttt{r}^2-\texttt{r}_-^2)}\,,\\
    e^1&=-\frac{\texttt{r} d\texttt{r}}{\xi \tau \sqrt{(\texttt{r}^2-\texttt{r}_+^2)(\texttt{r}^2-\texttt{r}_-^2)}}\,,\nonumber\\
    e^2&=-\frac{\xi  (\texttt{r}_++\texttt{r}_-)^2 (\tau d\texttt{t}  -d{\mathgtt{f}})-(\tau d\texttt{t}  +d{\mathgtt{f} }) \left(2 \texttt{r}^2-\texttt{r}_+^2-\texttt{r}_-^2\right)}{2 \xi  (\texttt{r}_++\texttt{r}_-)}\,.
\end{align}
The above expression is written in terms of new $(\texttt{t},\texttt{r},\mathgtt{f})$ coordinates, with $\texttt{r}_+\geq \texttt{r}_->0$. As in the previous case, $\texttt{r}_\pm$ are coordinates singularities for the radial coordinate, $\texttt{r}\in\mathbb R_+\setminus \{ \texttt{r}_\pm \}$, but, differently from before, the {extremal case} $\texttt{r}_+=\texttt{r}_-$ is allowed for any value of $\xi\in (0,1]$. In the range $\texttt{r}_-\leq\texttt{r}\leq \texttt{r}_+$, the vielbein can be made real by performing the complex transformation \eqref{complexLorentz}. The corresponding metric can again be written in fibration form
\begin{equation}
    ds^2=\left(g_{\texttt{tt}}-\frac{g_{\texttt{t}\mathgtt{f}}^2}{g_{\mathgtt{ff}}}\right)\,d\texttt{t}^2+g_{\texttt{rr}}\,d\texttt{r}^2+g_{\mathgtt{ff}}\left(d\mathgtt{f}+\frac{g_{\texttt{t}\mathgtt{f}}}{g_{\mathgtt{ff}}}\,d\texttt{t}\right)^2\,,
\end{equation}
with 
\begin{align}
g_{\texttt{rr}}&=-\frac{\texttt{r}^2}{(\texttt{r}^2-\texttt{r}_+^2)(\texttt{r}^2-\texttt{r}_-^2)\tau^2\xi^2}\,,\nonumber\\
g_{\texttt{tt}}-\frac{g_{\texttt{t}\mathgtt{f}}^2}{g_{\mathgtt{ff}}}&=\frac{(\texttt{r}^2-\texttt{r}_+^2)(\texttt{r}^2-\texttt{r}_-^2)}{\mathcal P(\texttt{r},\xi)}\,,\nonumber\\
g_{\mathgtt{ff}}&=-\tau^2\mathcal P(\texttt{r},\xi)\,,\nonumber\\
g_{\texttt{t}\mathgtt{f}}&=\tau \texttt{r}_+ \texttt{r}_--\frac{  \left(2 \texttt{r}^2-\texttt{r}_+^2-\texttt{r}_-^2\right)^2\left(1-\xi ^2\right) \tau}{4(\texttt{r}_++\texttt{r}_-)^2\xi^2}
\end{align}
and 
\begin{align}
    \mathcal P(\texttt{r},\xi)=\frac{\left(2 \texttt{r}^2-\texttt{r}_+^2-\texttt{r}_-^2+\xi  (\texttt{r}_++\texttt{r}_-)^2\right)^2-4 \xi ^2 \left(\texttt{r}^2-\texttt{r}_+^2\right)\left(\texttt{r}^2-\texttt{r}_-^2\right) }{4 \tau^2\xi ^2 (\texttt{r}_++\texttt{r}_-)^2}\,.
\end{align}
As in the previous case, the expression of the polynomial $\mathcal P(\texttt{r},\xi)$ greatly simplifies in the undeformed case
$$\lim_{\xi\to1}\mathcal P(\texttt{r},\xi)=\frac{\texttt{r}^2}{\tau^2}\,,$$
where we retrieve the AdS$_3$ spacetime.
The geometry admits two Killing vectors, which read
\begin{align}
    K_{\texttt{t}}=\frac{\partial}{\partial \texttt{t}}\,, \qquad K_{\mathgtt{f}}=\frac{\partial}{\partial \mathgtt{f}}\,,
\end{align}
whose norm can again be studied in relation to special kinds of hypersurfaces.\\ \newline
Let us repeat the discussion about the presence of Killing horizons and let us consider the null hypersurfaces generated by $f_\pm(\texttt{r})=\texttt{r}-\texttt{r}_\pm$. The following linear combinations become null (and mutually orthogonal) on each hypersurface respectively:
\begin{align}
    K'_\pm=\pm\frac{ \texttt{r}_+(1\pm\xi)- \texttt{r}_-(1\mp\xi)}{2\tau}K_{\texttt{t}}\mp\frac{ \texttt{r}_+(1\mp\xi)- \texttt{r}_-(1\pm\xi)}{2}K_{\mathgtt{f}}\,.
\end{align}
This indicates that $f_\pm(\texttt{r})$ define Killing hypersurfaces.\\ \newline
Before considering the presence of singularities, let us describe the ergospheres in terms of the zeros of $g_{\texttt{tt}}(\texttt{r})$. The latter explicitly reads
\begin{align}
    g_{\texttt{tt}}(\texttt{r})=\frac{\tau ^2}{(\texttt{r}_++\texttt{r}_-)^2}{  \left(\left(\texttt{r}^2-\texttt{r}_+^2\right)\left(\texttt{r}^2-\texttt{r}_-^2\right)-\frac{\left(2 \texttt{r}^2-\texttt{r}_+^2-\texttt{r}_-^2-\xi  (\texttt{r}_++\texttt{r}_-)^2\right)^2}{4 \xi ^2}\right)}{}\,.
\end{align}
Its zeros are real and positive provided that 
\begin{align}
    \mathgtt{x}^*\leq \xi <1\,,\qquad \mathgtt{x}^*=\frac{\texttt{r}_+-\texttt{r}_-}{\sqrt{2(\texttt{r}_+^2+\texttt{r}_-^2)}}
\end{align}
and are given by
\begin{align}\label{zerosextremalcase}
    \texttt{r}^*_\pm=\sqrt{\frac{\left(1-\xi ^2\right) \left(\texttt{r}_+^2+\texttt{r}_-^2\right)+\xi  (\texttt{r}_++\texttt{r}_-)^2\pm\xi  (\texttt{r}_+^2-\texttt{r}_-^2)\sqrt{\frac{\xi^2}{\mathgtt{x}^{*2}} -1}}{2 \left(1-\xi ^2\right)}}\,.
\end{align}
As in the previous case, for any $\xi$ in this range, we have a set of inequalities 
\begin{align}
    0<\texttt{r}_-\leq \texttt{r}_+\leq \texttt{r}^*_{-}\leq \texttt{r}^*_{+}\,.
\end{align}
The last equality holds at $\xi=\mathgtt{x}^*$, at which value we have
\begin{align}\label{rpiumeno}
    \texttt{r}^*_+=\texttt{r}^*_-=\sqrt{(\texttt{r}_+^2+\texttt{r}_-^2){\left(\mathgtt{x}^*+\frac12\right)}}\,.
\end{align}
Let us observe that a bit of care is needed in the extremal limit, where $\mathgtt{x}^*=0$: this value is outside our range of validity, $(0,1)$. This simply indicates that the zeros \eqref{zerosextremalcase} are always real and positive. The corresponding surfaces, however, will get closer and closer to each other and to $\texttt{r}_+=\texttt{r}_-$, as $\xi$ decreases, without actually coinciding.\\
In general, contrary to the previous analysis, the value \eqref{rpiumeno} is not necessarily larger than the BTZ ergosphere radius, $\sqrt{\texttt{r}_+^2+\texttt{r}_-^2}$, and in particular we have another possible equality, $\texttt{r}_+=\texttt{r}^*_-$, corresponding to the value
\begin{align}
    \tilde\xi=\frac{\texttt{r}_+-\texttt{r}_-}{\texttt{r}_++\texttt{r}_-}> \mathgtt{x}^* \,.
\end{align}
{Again, the extremal case requires some attention, as $\tilde\xi=0$ is again excluded and cannot be reached.}\\
In the undeformed limit $\xi\to1$, the zeros \eqref{zerosextremalcase} coincide with the boundary at infinity and with the proper ergosphere, as in \eqref{undeflimit}. The behaviour of $g_{\texttt{tt}}$ and its zeros are shown in Figure \ref{gttExtr} and \ref{zeroExtr}.
\begin{figure}[ht]
\centering
\includegraphics[width=0.7\textwidth]{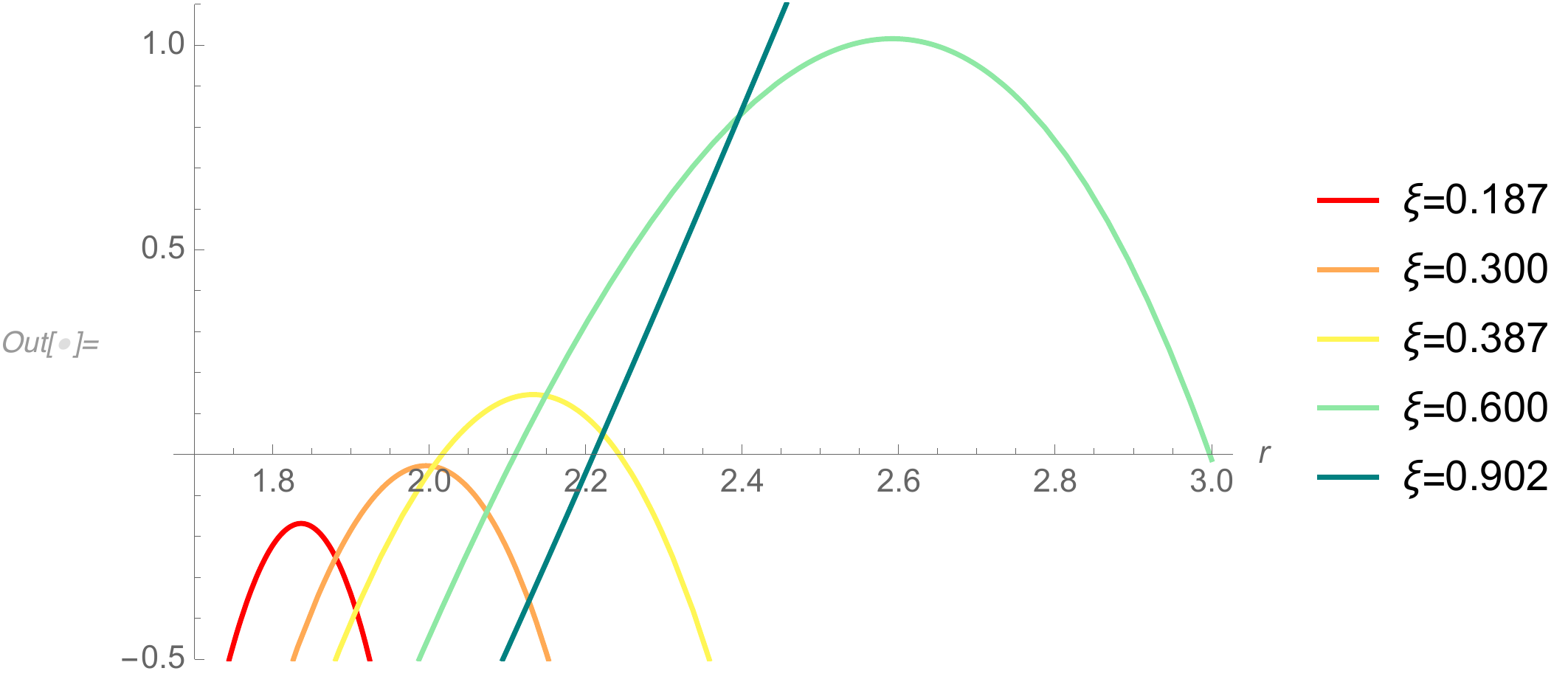}
\caption{The behaviour of $g_{\texttt{tt}}(r)$ is shown here for $\texttt{r}_-=1$, $\texttt{r}_+=2$ and for different values of $\xi$. In this case $\mathgtt{x}^*\simeq0,316$, which is consistent with the zeros that we see appearing starting from the yellow line.}\label{gttExtr}
\end{figure}
\begin{figure}[ht]
\centering
\includegraphics[width=.435\textwidth]{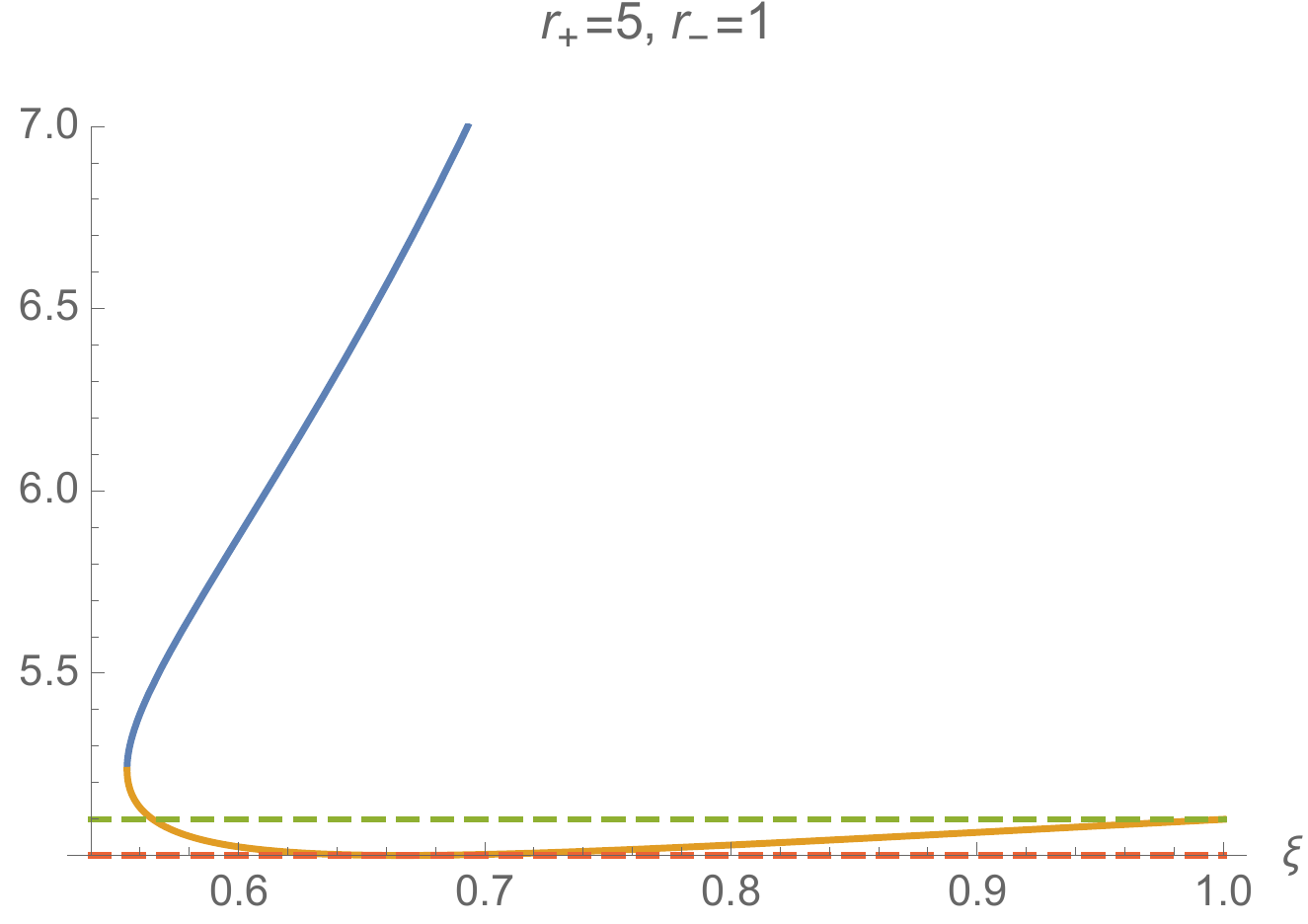}\includegraphics[width=.55\textwidth]{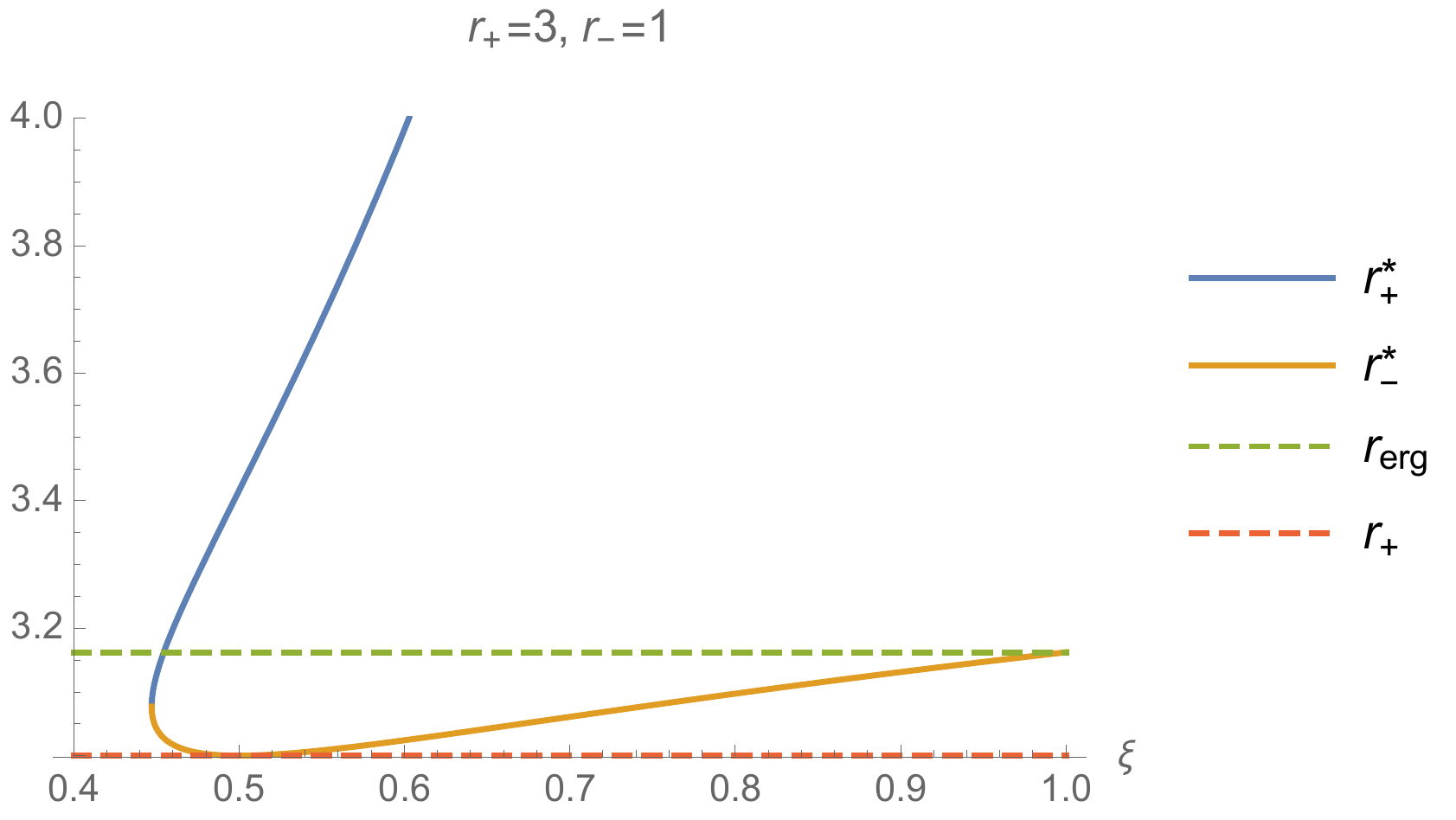}
\caption{The behaviour of the zeros of $g_{\texttt{tt}}(r)$ is shown here for different values of $\texttt{r}_+$ and $\texttt{r}_-$. In both cases $\texttt{r}^*_-$ touches $\texttt{r}_+$ before reaching the BTZ ergosphere value. The value of $\xi$ corresponding to the point where the blue and orange lines touch is $\mathgtt{x}^*$.}\label{zeroExtr}
\end{figure}
For small deformations, $\xi>\mathgtt{x}^*$, the solution admits a region where the norm of the Killing vector $K_{\texttt{t}}$ is positive, $\texttt{r}^*_-<\texttt{r}<\texttt{r}^*_+$, indicating the presence of static, non-rotating observers between the two ergospheres. This region shrinks to zero below the value $\mathgtt{x}^*$, as the zeros of $g_{\texttt{tt}}$ cease to exist.\\ \newline
At last, let us identify points with respect to the Killing vector $K_{\mathgtt f}$: this is again performed by identifying points such that $\mathgtt{f}\sim\mathgtt{f}+2\pi$, thus making $\mathgtt{f}$ compact. This identification is only consistent if we do not introduce closed timelike/lightlike curves. This in turn depends on the norm of $K_{\mathgtt f}$ and ultimately on the zeros of the metric component $g_{\mathgtt{ff}}(r)$. Differently from the previously analysed case, the polynomial $\mathcal P(\texttt{r},\xi)$ may admit zeros of the following form: 
\begin{align}
    \hat r_\pm=\sqrt{\frac{\left(1-\xi ^2\right) (\texttt{r}_+^2+\texttt{r}_-^2)-\xi  (\texttt{r}_++\texttt{r}_-)^2\pm\xi  \left(\texttt{r}_+^2-\texttt{r}_-^2\right) \sqrt{\frac{\xi^2}{\mathgtt{x}^{*2}}-1}}{2 \left(1-\xi ^2\right)}}\,.
\end{align}
By requiring the above expressions to be real, one can derive several conditions involving $\xi$, $\texttt{r}_-$ and $\texttt{r}_+$. However, independently of the latter relations, we always find that $\hat r_-\leq\hat r_+\leq\texttt{r}_-$, where the last equality is reached at $\xi=\tilde\xi$. These causal singularities are therefore protected by the Killing horizons $\texttt{r}_\pm$, which can be interpreted as event horizons. Moreover, in the undeformed limit, we only have one singularity, which precisely coincides with the BTZ one,
\begin{align}
    \lim_{\xi\to1}\hat r_+=0\,.
\end{align}
At bit of care is needed again in the extremal case, $\texttt{r}_+=\texttt{r}_-=r_0$, where the zeros reduce to 
\begin{align}
    \hat r_+=r_0\sqrt\frac{1-\xi}{1+\xi}\,,\qquad \hat r_-=r_0\sqrt\frac{1-3\xi}{1-\xi}\,,
\end{align}
the latter only existing for $0<\xi\leq1/3$. Since the parameter $\xi$ is strictly positive, the singularities never reach $r_0$, by decreasing $\xi$, and therefore never actually coincide with the ergospheres. In Figure \ref{sing} we can see an example of the behaviour of ergospheres and causal singularities in both non-extremal and extremal cases.
\begin{figure}[ht]
\centering
\includegraphics[width=.45\textwidth]{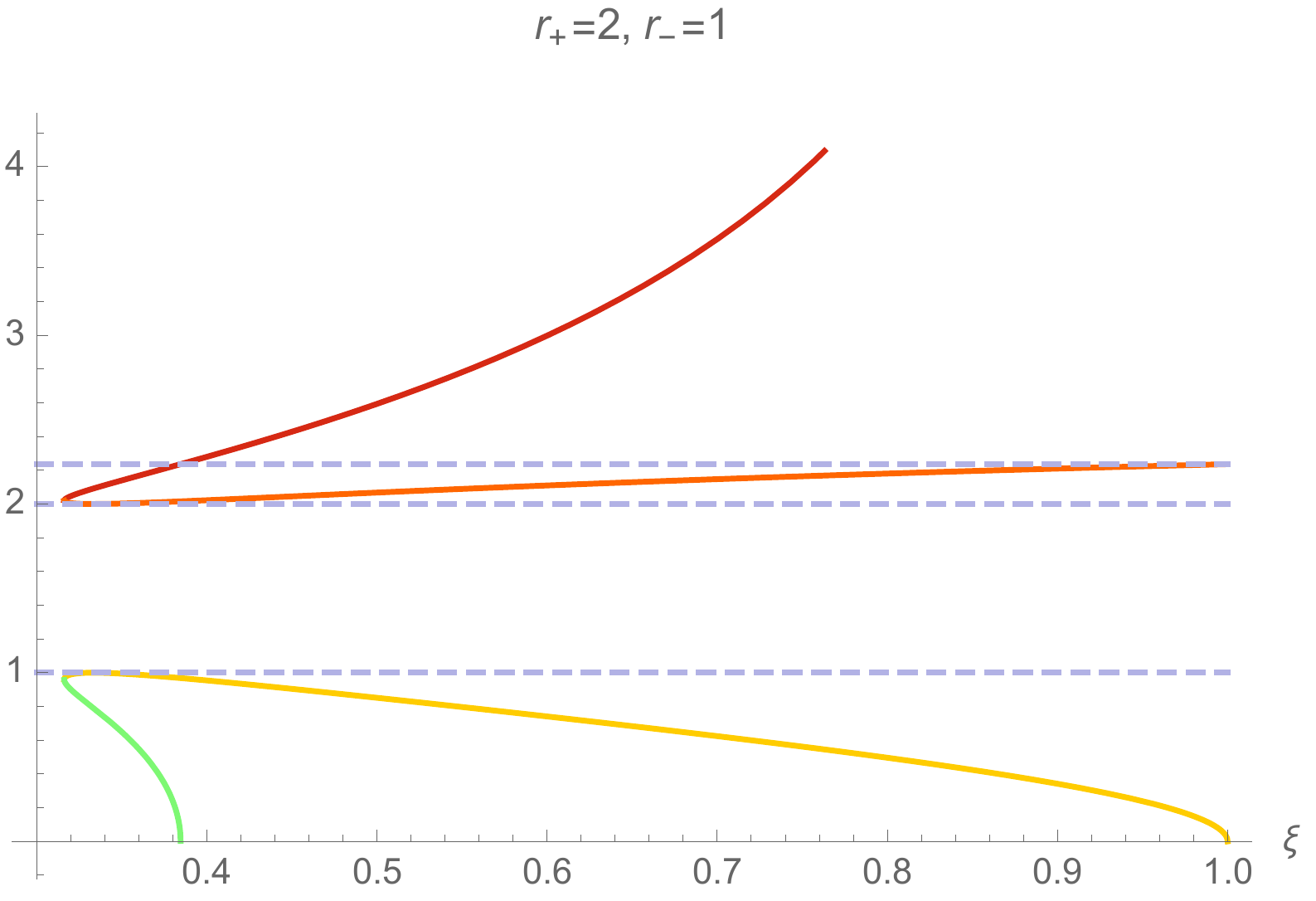}\includegraphics[width=.53\textwidth]{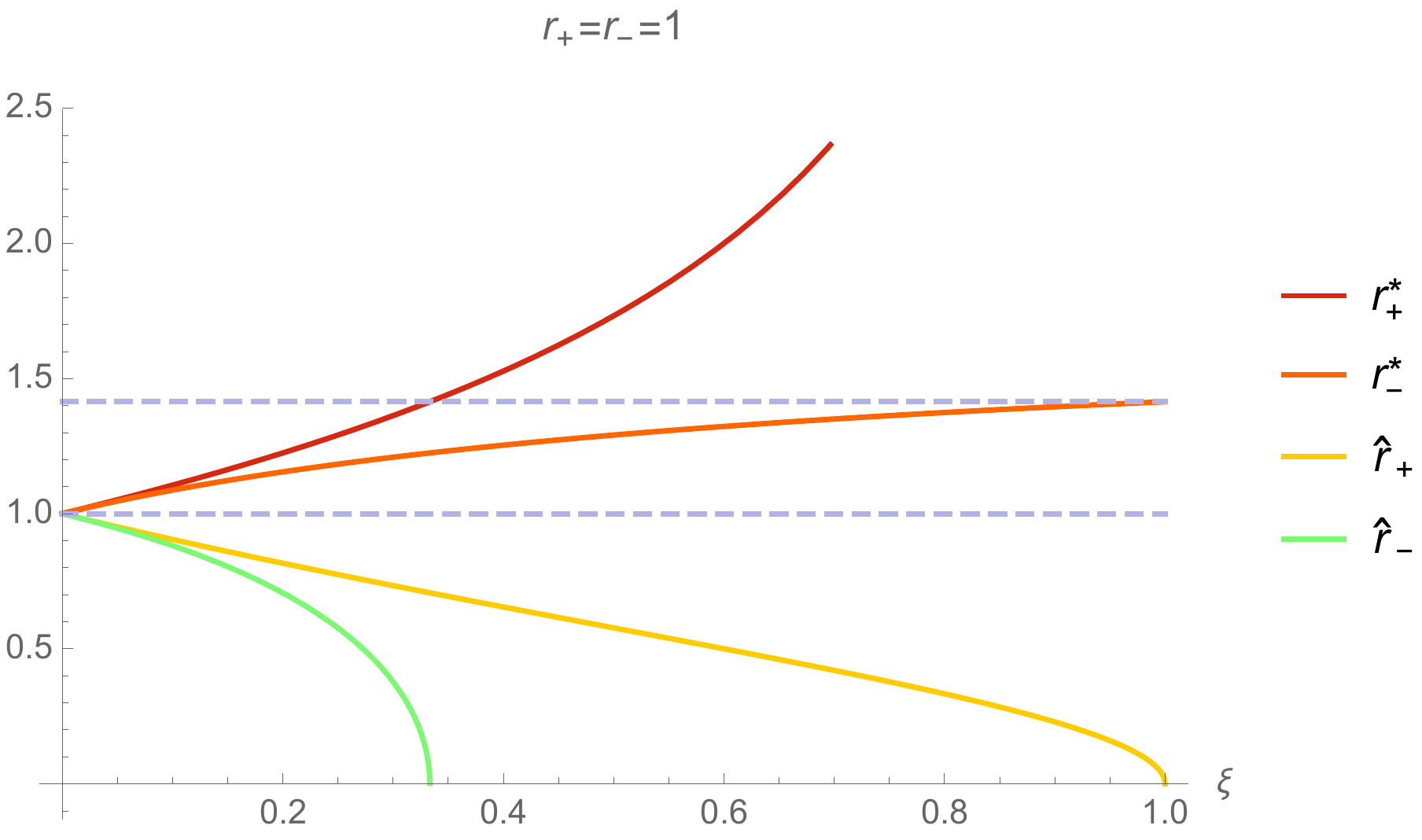}
\caption{On the left, we represent ergospheres and singularities in the non-extremal case, whereas on the right we have the extremal one. The dashed lines represent (from top to bottom) $r_{\rm erg}$, $\texttt{r}_+$ and $\texttt{r}_-$.}\label{sing}
\end{figure}\\
The fact that the causal structures of the two spacetimes we described differ should not surprise. Their physical inequivalence is motivated by the fact that they were obtained by quotienting the same covering spacetime with respect to two different discrete groups of motion, generated by $K_\phi$ and $K_{\mathgtt f}$, respectively.

\subsection{Timelike fibration}
A different solution to equations \eqref{R} and \eqref{T} can be found by considering a constant totally antisymmetric torsion component and a trace part proportional to $e^0$, in such a way that the Beltrami torsion is timelike. In this case, one explicitly gets
\begin{align}\nonumber
    e^0&=\frac{1}{2 \tau}\left({{dt}+{dx} \left(A\frac{ \sinh \left(\sqrt{\zeta ^2+1} z\right)}{\sqrt{\zeta ^2+1}}+B\frac{\cosh \left(\sqrt{\zeta ^2+1} z\right)}{\sqrt{\zeta ^2+1}}\right)}\right)\,,\\ \nonumber
    e^1&=\frac{1}{2\tau}{{dx} \left(A \cosh \left(\sqrt{\zeta ^2+1} z\right)+B \sinh \left(\sqrt{\zeta ^2+1} z\right)\right)}\,,\\ \nonumber
    e^2&=\frac{1}{2\tau}dz\,,\\ 
    \beta&=2\tau\zeta e^0\,,
\end{align}
which has been written in terms of local coordinates $(t,x,z)$ each covering the entire real line. Differently from the spacelike fibration case, we have here two real parameters, $A$ and $B$, which cannot simply be eliminated by a real change of coordinates. \\
The corresponding metric admits isometries generated by linear combinations of the following two Killing vectors:
\begin{align}
    K_t=\frac{\partial}{\partial t}\,,\qquad K_x=\frac{\partial}{\partial x}\,.
\end{align}
The torsionless Riemann tensor reads
\begin{align}
    R^{01}[\mathring\omega]&=\tau^2 e^0\wedge e^1\,, \qquad  R^{02}[\mathring\omega]=\tau^2 e^0\wedge e^2\,,\qquad  R^{12}[\mathring\omega]=(1+4\zeta^2)\tau^2 e^1\wedge e^2\,,
\end{align}
while the Ricci scalar is
\begin{align}
    R[\mathring\omega]&=6\tau^2+8\zeta^2\tau^2\,.
\end{align}
We again notice that $\zeta\in\mathbb R$ parametrises the deformation with respect to AdS$_3$: the trace component of the torsion, $\beta$, introduces a squashing along the $12$ direction, while maintaining the structure of an Einstein manifold.\\ \newline
The energy-momentum tensor associated to the chosen torsion is
\begin{align}
    \mathcal T^{ij}=\left(-\tau^2\eta^{ij}-4\tau^2\zeta^2\delta^i_0\delta^j_0\right)\,,
\end{align}
which is evidently symmetric and reduces to a cosmological constant contribution in the vanishing $\zeta$ limit. 

\subsection{Solutions for non-relativistic Beltrami fluids} \label{euclidsol}
Let us consider the spacelike fibration solution \eqref{sol} and perform a complex SL$(3,\mathbb C)$ transformation $e^i_E\equiv \mathcal S^i{}_{j}e^j$, with
\begin{align}
    \mathcal S=\begin{pmatrix} -1& 0& 0 \\
    0&i&0\\
    0&0&i\end{pmatrix}\,,
\end{align} 
in combination with $\zeta\to -i\zeta$, in order to keep $\beta$ real. In particular, the complex transformation $\mathcal S$ has been chosen in such a way as to transform the SO$(1,2)$ invariant tensor $\eta_{ij}$ into the SO$(3)$ one, $\delta_{ij}$, meaning that the signature turns Euclidean. The full solution reads
\begin{align}\nonumber
    e^0_E&=-\frac{1}{2 \tau}dt\,,\\ \nonumber    
    e^1_E&=\frac{1}{2\tau}{{dx} \left( \cos \left(\sqrt{1+\zeta^2} t\right)\right)}\,,\\ \nonumber
    e^2_E&=\frac{1}{2\tau}\left({dz}+dx\left({\,\frac{  \sin \left(\sqrt{1+\zeta^2} t\right)}{\sqrt{1+\zeta^2}}}\right)\right)\,,\\
    \beta&=2\tau\zeta e^2_E\,. 
\end{align}
Since all coordinates are on equal footing in an Euclidean space, the deformations of $S^3$ along the other directions can be obtained by means of SO$(3)$ rotations
\begin{align}
    \mathcal R_{12}=\begin{pmatrix}  0&0& 1 \\
    0&1&0\\
    -1&0&0\end{pmatrix}\,,\qquad \mathcal R_{02}=\begin{pmatrix}  1&0& 0 \\
    0&0&1\\
    0&-1&0\end{pmatrix}\,.
\end{align}
In the Euclidean case, the integrability equation \eqref{ArnoldBeltrami} coincides with the Beltrami equation, describing the velocity of non-relativistic incompressible three-dimensional Beltrami fluids. In this context, the velocity of the fluid is precisely the vector field $\beta_\mu$, whose interpretation in terms of four-dimensional energy-momentum tensor will be discussed in a forthcoming publication.

\section{Conclusions}\label{sec4}
One of the goals of the present paper is to emphasise the role of the spacetime torsion, in the Einstein-Cartan gravity formalism, as a natural, geometric way to couple gravity to matter.  \\
The integrability conditions on the considered irreducible components of the torsion, $\tau$ and $\beta_i$, which follow from the nilpotency of the differential operator, guarantee the consistent coupling to gravity of this particular matter source, i.e. the covariant conservation of the corresponding energy-momentum tensor. We found, to our surprise, that these integrability conditions allow for configurations, which were not considered in the literature, featuring a component $\beta_i$ of the torsion, which, seen as a 1-form $\beta=\beta_i e^i$, is not closed.
In fact it satisfies a dynamical equation reminiscent of the one describing Beltrami fluids. 
Since $\beta$ has the additional meaning of a connection associated with local Weyl rescalings, the corresponding O$(1,1)$-bundle is curved. \\
We then presented different families of solutions to the curvature and torsion equations \eqref{R},\eqref{T}. They depend on a continuous parameter, which measures the deviation from a  locally AdS$_3$ geometry, induced by the Beltrami torsion $\beta$. The latter can be interpreted as a geometric flux, capable of removing, in certain cases, the usual causal singularity of the BTZ black hole. \\
Moreover, as we discussed in the main text, we found that, in the spacelike fibration case, the Beltrami torsion breaks the conformal geometry of the two-dimensional boundary. This peculiar feature is worth investigating within the gauge/gravity correspondence. \\
Lastly, the Euclidean solutions discussed in Section \ref{euclidsol} provide concrete and novel configurations for the so-called Beltrami fluids, on non-trivial curved spaces. It would be interesting to embed these solutions in a four-dimensional spacetime where the Beltrami torsion could be interpreted as the velocity field of a Beltrami fluid, and study, in this setting, the deformations induced by such fluid on the geometry of the four-dimensional spacetime.\\ \newline
In this paper, besides deriving these new three-dimensional solutions, we have presented a first analysis of their geometry. We leave to a future investigation a thorough description of the causal structure of these spacetimes and of their possible applications to the gauge/gravity correspondence. In particular, it would now be interesting to derive the surface gravity associated with the Killing horizons and the related black hole temperature, in those cases in which a singularity exists.\\
The setup discussed in the present paper could be generalised in several directions: one would be to include, as a possible source of gravitational matter, the symmetric traceless irreducible component of the torsion $\Sigma^{ij}$, which we disregarded in the present paper, and look for consistent solutions of the relevant equations. 
It would also be interesting to include dynamical matter, such as spin-$1/2$ fermion fields. This is the case with Unconventional supersymmetry models, which are also appealing because they are based on a Weyl-invariant action. This particular application is under investigation and will be the subject of a forthcoming publication.

\section*{Acknowledgments}
We wish to thank R. D'Auria and P. Fr{\`e} for their useful comments. R. N. would like to thank Carlo Alberto Cremonini and Joris Raeymaekers for useful discussions and acknowledges GAČR grant EXPRO 20-25775X for financial support. L. R. would like to thank the Department of Applied Science and Technology of the Polytechnic of Turin for financial support. J. Z. has been partially supported by FONDECYT/ANID grant 1220862.

\appendix

\section{Conventions}\label{appA}
The three-dimensional index conventions are the following
\begin{align}\nonumber
    i,j&=0,1,2\qquad \text{Lorentz rigid indices,}\\
    \mu,\nu&=0,1,2\qquad \text{Spacetime curved indices.}
\end{align}
A differential $p$-form ($p\leq3$) can be expressed as
\begin{align}
    \omega^{(p)}=\frac{1}{p!}\,\omega_{\mu_1\ldots \mu_p}dx^{\mu_1}\wedge\ldots\wedge dx^{\mu_p}\,,
\end{align}
where $\wedge$ is the wedge-product.\\ 
The signature of the invariant tensor of the Lorentz group SO$(1,2)$ is $(+,-,-)$, whereas the Levi-Civita symbol is 
\begin{align}
    \epsilon_{012}=\epsilon^{012}=1\,.
\end{align}
Finally, the Hodge dual of a $p$-form is defined as
\begin{align}
    \star\,\omega^{(p)}=\frac{\sqrt{g}}{p!(d-p)!}\epsilon_{\mu_1\ldots\mu_d}\,\omega^{\mu_1\ldots\mu_p}\,dx^{\mu_{p+1}}\wedge\ldots\wedge dx^{\mu_d}\,,
\end{align}
where $d=3$.

\section{Isometries and Killing vectors of \texorpdfstring{AdS$_3$}{AdS3}%
}\label{appB}
The $\mathfrak{so}(2,2)$ Killing vectors of AdS$_3$ in the local coordinate patch used in \eqref{metriccomp} read
\begin{align}\nonumber
    K_{01}&=\frac{ r_-}{\tau(r_+^2-r_-^2)}\frac{\partial}{\partial\text{t}}+\frac{r_+}{r_+^2-r_-^2}\frac{\partial}{\partial\phi}\,,\\ \nonumber
    K_{02}&=-\frac{g_+(r) \sinh (m_+(r))+g_-(r)\sinh (m_-(r))}{2 \tau\sqrt{\left(r^2-r_+^2\right)\left(r^2-r_-^2\right) }}\frac{\partial}{\partial\text{t}}\\ \nonumber
    &+\sqrt{\left(r^2-r_+^2\right)\left(r^2-r_-^2\right)}\,\frac{  \cosh (m_+(r))+\cosh (m_-(r))}{2 r}\frac{\partial}{\partial r}\\ \nonumber
    &+\frac{h_+(r)\sinh (m_+(r))-h_-(r)\sinh (m_-(r))}{2 \sqrt{ \left(r^2-r_+^2\right)\left(r^2-r_-^2\right)}}\frac{\partial}{\partial \phi}\,,\\ \nonumber
    K_{03}&=\frac{ g_+(r)\cosh (m_+(r))+g_-(r)\cosh (m_-(r))}{2 \tau  \sqrt{\left(r^2-r_+^2\right)\left(r^2-r_-^2\right) }}\frac{\partial}{\partial \text{t}} \\\nonumber
    &-\sqrt{\left(r^2-r_+^2\right)\left(r^2-r_-^2\right) }\,\frac{ \sinh (m_+(r))+\sinh (m_-(r))}{2 r}\frac{\partial}{\partial r} \\ \nonumber
    &-\frac{h_+(r)\cosh (m_+(r))-h_-(r)\cosh (m_-(r))}{2 \sqrt{\left(r^2-r_+^2\right)\left(r^2-r_-^2\right) }}\frac{\partial}{\partial \phi}\,,\\ \nonumber 
    K_{12}&=-\frac{ g_+(r)\cosh (m_+(r))-g_-(r)\cosh (m_-(r))}{2 \tau  \sqrt{\left(r^2-r_+^2\right)\left(r^2-r_-^2\right) }}\frac{\partial}{\partial \text{t}}\\ \nonumber 
    &+\sqrt{\left(r^2-r_+^2\right)\left(r^2-r_-^2\right) }\,\frac{  (\sinh (m_+(r))-\sinh (m_-(r)))}{2 r}\frac{\partial}{\partial r}\\ \nonumber
    &+\frac{ h_+(r)\cosh (m_+(r))+h_-(r)\cosh (m_-(r))}{2 \sqrt{\left(r^2-r_+^2\right)\left(r^2-r_-^2\right) }}\frac{\partial}{\partial \phi}\,,\\ \nonumber
    K_{13}&=\frac{g_+(r)\sinh (m_+(r))-g_-(r)\sinh (m_-(r))}{2 \tau  \sqrt{\left(r^2-r_+^2\right)\left(r^2-r_-^2\right) }}\frac{\partial}{\partial \text{t}}\\ \nonumber 
    &-\sqrt{ \left(r^2-r_+^2\right)\left(r^2-r_-^2\right)}\,\frac{  \cosh (m_+(r))-\cosh (m_-(r))}{2 r}\frac{\partial}{\partial r}\\ \nonumber
    &-\frac{ h_+(r)\sinh (m_+(r))+h_-(r)\sinh (m_-(r))}{2 \sqrt{ \left(r^2-r_+^2\right)\left(r^2-r_-^2\right)}}\frac{\partial}{\partial \phi}\,,\\ 
    K_{23}&=-\frac{ r_+}{\tau(r_+^2-r_-^2)}\frac{\partial}{\partial\text{t}}-\frac{r_-}{r_+^2-r_-^2}\frac{\partial}{\partial\phi}\,,
\end{align}
where we defined 
\begin{align}\nonumber
    g_{\pm}(r)=\frac{r^2\pm r_+r_-}{r_+\pm r_-}\,, \quad h_\pm(r)=\frac{r^2-r_+^2\mp r_+r_--r_-^2}{r_+\pm r_-}\,, \quad m_\pm=(r_+\pm r_-)(\tau \text{t}\mp\phi)\,.
\end{align}
{These vectors satisfy the following Lie algebra relation
\begin{align}
    [K_{ab},K_{cd}]=K_{ad}\eta^{(4)}_{bc}+K_{bc}\eta^{(4)}_{ad}-K_{ac}\eta^{(4)}_{bd}-K_{bd}\eta^{(4)}_{ac}\,,
\end{align}
where $a,b=0,\ldots,3$ and $\eta^{(4)}$ is the SO$(2,2)$ invariant tensor.}\\
The solution \eqref{metriccomp} only preserves a two-dimensional $\mathfrak{so}(2)\oplus\mathfrak{so}(2)$ subalgebra of the $\mathfrak{so}(2,2)$ one, generated by 
\begin{align}
    K_{\text{t}}=-\tau\left(r_-K_{01}+r_+K_{23}\right)\,,\qquad K_{\phi}=r_+K_{01}+r_-K_{23}\,,
\end{align}
with $[K_{\text t},K_{\phi}]=0$.



\begin{thebibliography}{99}

\bibitem{Brown:1986nw}
J.~D.~Brown and M.~Henneaux,
``Central Charges in the Canonical Realization of Asymptotic Symmetries: An Example from Three-Dimensional Gravity,''
Commun. Math. Phys. \textbf{104} (1986), 207-226
doi:10.1007/BF01211590

\bibitem{Achucarro:1986uwr}
A.~Achucarro and P.~K.~Townsend,
``A Chern-Simons Action for Three-Dimensional anti-De Sitter Supergravity Theories,''
Phys. Lett. B \textbf{180} (1986), 89
doi:10.1016/0370-2693(86)90140-1

\bibitem{Witten:1988hc}
E.~Witten,
``(2+1)-Dimensional Gravity as an Exactly Soluble System,''
Nucl. Phys. B \textbf{311} (1988), 46
doi:10.1016/0550-3213(88)90143-5

\bibitem{Achucarro:1989gm}
A.~Achucarro and P.~K.~Townsend,
``Extended Supergravities in $d$ = (2+1) as {Chern-Simons} Theories,''
Phys. Lett. B \textbf{229} (1989), 383-387
doi:10.1016/0370-2693(89)90423-1

\bibitem{Coussaert:1995zp}
O.~Coussaert, M.~Henneaux and P.~van Driel,
``The Asymptotic dynamics of three-dimensional Einstein gravity with a negative cosmological constant,''
Class. Quant. Grav. \textbf{12} (1995), 2961-2966
doi:10.1088/0264-9381/12/12/012
[arXiv:gr-qc/9506019 [gr-qc]].

\bibitem{Witten:2007kt}
E.~Witten,
``Three-Dimensional Gravity Revisited,''
[arXiv:0706.3359 [hep-th]].

\bibitem{Banados:1992wn}
M.~Banados, C.~Teitelboim and J.~Zanelli,
``The Black hole in three-dimensional space-time,''
Phys. Rev. Lett. \textbf{69} (1992), 1849-1851
doi:10.1103/PhysRevLett.69.1849
[arXiv:hep-th/9204099 [hep-th]].

\bibitem{Banados:1992gq}
M.~Banados, M.~Henneaux, C.~Teitelboim and J.~Zanelli,
``Geometry of the (2+1) black hole,''
Phys. Rev. D \textbf{48} (1993), 1506-1525
[erratum: Phys. Rev. D \textbf{88} (2013), 069902]
doi:10.1103/PhysRevD.48.1506
[arXiv:gr-qc/9302012 [gr-qc]].

\bibitem{Anninos:2008fx}
D.~Anninos, W.~Li, M.~Padi, W.~Song and A.~Strominger,
``Warped AdS(3) Black Holes,''
JHEP \textbf{03} (2009), 130
doi:10.1088/1126-6708/2009/03/130
[arXiv:0807.3040 [hep-th]].

\bibitem{Maldacena:1997re}
J.~M.~Maldacena,
``The Large N limit of superconformal field theories and supergravity,''
Adv. Theor. Math. Phys. \textbf{2} (1998), 231-252
doi:10.4310/ATMP.1998.v2.n2.a1
[arXiv:hep-th/9711200 [hep-th]].

\bibitem{Ryu:2006bv}
S.~Ryu and T.~Takayanagi,
``Holographic derivation of entanglement entropy from AdS/CFT,''
Phys. Rev. Lett. \textbf{96} (2006), 181602
doi:10.1103/PhysRevLett.96.181602
[arXiv:hep-th/0603001 [hep-th]].

\bibitem{Nishioka:2009un}
T.~Nishioka, S.~Ryu and T.~Takayanagi,
``Holographic Entanglement Entropy: An Overview,''
J. Phys. A \textbf{42} (2009), 504008
doi:10.1088/1751-8113/42/50/504008
[arXiv:0905.0932 [hep-th]].

\bibitem{deBoer:2013vca}
J.~de Boer and J.~I.~Jottar,
``Entanglement Entropy and Higher Spin Holography in AdS$_3$,''
JHEP \textbf{04} (2014), 089
doi:10.1007/JHEP04(2014)089
[arXiv:1306.4347 [hep-th]].

\bibitem{Alvarez:2014uda}
P.~D.~Alvarez, P.~Pais, E.~Rodr\'\i{}guez, P.~Salgado-Rebolledo, and J.~Zanelli, ``The BTZ black hole as a Lorentz-flat geometry'', Phys. Lett. B \textbf{738} (2014), 134
doi:10.1016/j.physletb.2014.09.032

\bibitem{Alvarez:2011gd}
P.~D.~Alvarez, M.~Valenzuela and J.~Zanelli,
``Supersymmetry of a different kind,''
JHEP \textbf{04} (2012), 058
doi:10.1007/JHEP04(2012)058
[arXiv:1109.3944 [hep-th]].

\bibitem{ABC3}
E.~Beltrami, ``Opera matematiche'', \textbf{4} (1889) 304.

\bibitem{Fre:2022odf}
P.~G.~Fr\'e and M.~Trigiante,
``Chaos from Symmetry: Navier Stokes equations, Beltrami fields and the Universal Classifying Crystallographic Group,''
[arXiv:2204.01037 [nlin.CD]].

\bibitem{Alvarez:2013tga}
P.~D.~Alvarez, P.~Pais and J.~Zanelli,
``Unconventional supersymmetry and its breaking,''
Phys. Lett. B \textbf{735} (2014), 314-321
doi:10.1016/j.physletb.2014.06.031
[arXiv:1306.1247 [hep-th]].

\bibitem{Guevara:2016rbl}
A.~Guevara, P.~Pais and J.~Zanelli,
``Dynamical Contents of Unconventional Supersymmetry,''
JHEP \textbf{08} (2016), 085
doi:10.1007/JHEP08(2016)085
[arXiv:1606.05239 [hep-th]].

\bibitem{Andrianopoli:2018ymh}
L.~Andrianopoli, B.~L.~Cerchiai, R.~D'Auria and M.~Trigiante,
``Unconventional supersymmetry at the boundary of AdS$_{4}$ supergravity,''
JHEP \textbf{04} (2018), 007
doi:10.1007/JHEP04(2018)007
[arXiv:1801.08081 [hep-th]].

\bibitem{Andrianopoli:2019sip}
L.~Andrianopoli, B.~L.~Cerchiai, R.~D'Auria, A.~Gallerati, R.~Noris, M.~Trigiante and J.~Zanelli,
``$\mathcal{N}$-extended $D = 4$ supergravity, unconventional SUSY and graphene,''
JHEP \textbf{01} (2020), 084
doi:10.1007/JHEP01(2020)084
[arXiv:1910.03508 [hep-th]].

\bibitem{Alvarez:2021zhh}
P.~D.~Alvarez, L.~Delage, M.~Valenzuela and J.~Zanelli,
``Unconventional SUSY and Conventional Physics: A Pedagogical Review,''
Symmetry \textbf{13} (2021) no.4, 628
doi:10.3390/sym13040628
[arXiv:2104.05133 [hep-th]].

\bibitem{Jones:1985pla}
P.~Jones and K.~Tod,
``Minitwistor spaces and Einstein-Weyl spaces,''
Class. Quant. Grav. \textbf{2} (1985) no.4, 565-577
doi:10.1088/0264-9381/2/4/021

\bibitem{Klemm:2020gfm}
S.~Klemm and L.~Ravera,
``Schr\"odinger connection with selfdual nonmetricity vector in 2+1 dimensions,''
Phys. Lett. B \textbf{817} (2021), 136291
doi:10.1016/j.physletb.2021.136291
[arXiv:2008.12740 [hep-th]].





















\end{thebibliography}
\end{document}